\documentclass[%
 reprint,
superscriptaddress,
longbibliography,
 amsmath,amssymb,
 aps,prx,
floatfix,
]{revtex4-2}

\usepackage{graphicx}
\usepackage{dcolumn}
\usepackage{bm}
\usepackage{bbm}
\usepackage[colorlinks=true]{hyperref}
\usepackage{bbold}
\usepackage{makecell,hhline}
\usepackage{amsmath,mathtools,amssymb}
\usepackage{xcolor}
\usepackage{soul}
\usepackage{tikz}
\usepackage{physics}
\usetikzlibrary{decorations.pathreplacing}

\usepackage{ytableau}

\newcommand{\ra}{\rangle}

\newcommand{\ignore}[1]{{}}
\newcommand{\nobibentry}[1]{{\let\nocite\ignore\bibentry{#1}}}

\newcommand{\bibfnamefont}[1]{#1}
\newcommand{\bibnamefont}[1]{#1}

\newtheorem{res}{Result}

\begin{document}

\title{How exchange symmetry impacts performance of collective quantum heat engines
}

\author{Julia Boeyens}
 \email{julia.boeyens@gmail.com}
\affiliation{Naturwissenschaftlich-Technische Fakult\"at, Universit\"at Siegen, Walter-Flex-Strasse 3, 57068 Siegen, Germany}
\author{Benjamin Yadin}
\affiliation{Naturwissenschaftlich-Technische Fakult\"at, Universit\"at Siegen, Walter-Flex-Strasse 3, 57068 Siegen, Germany}
\author{Stefan Nimmrichter}
\affiliation{Naturwissenschaftlich-Technische Fakult\"at, Universit\"at Siegen, Walter-Flex-Strasse 3, 57068 Siegen, Germany}

\date{\today}

\begin{abstract}
Recently, multilevel collectively coupled quantum machines like heat engines and refrigerators have been shown to admit performance enhancements in analogy to superradiance. Thus far, investigations of the performance of collective quantum machines have largely restricted the dynamics to particles with bosonic exchange symmetry, especially for large numbers of particles. However, collections of indistiguishable but not fundamentally identical particles may assume quantum states of more general exchange symmetry or combinations thereof, raising the question of whether collective advantages can be observed for dynamics that allow the full Hilbert space to be explored.
Here, we compare a collection of single-particle three-level masers with their collectively coupled counterpart, while admitting more general forms of exchange symmetry. We study ergotropy and emitted power as the figures of merit and show which of the known results applicable to a single three-level engine carry over to an engine made up of a collectively coupled ensemble. We do this using results from representation theory to characterise the full basis of the Hilbert space and provide general tools for the description of the dynamics of such systems.
We find that collective work extraction can extend beyond the temperature window of three-level lasing, whereas in the lasing regime, individual may outperform collective operation.
In addition, the optimal parameter regime for work-like energy output varies for different symmetry types. 
Our results show a rich picture in which bosonic symmetry is not always optimal and sometimes individual particles may even perform best. 
\end{abstract}

\maketitle

\section{Introduction}
\begin{figure}
\centering
   \includegraphics[width=0.35\textwidth]{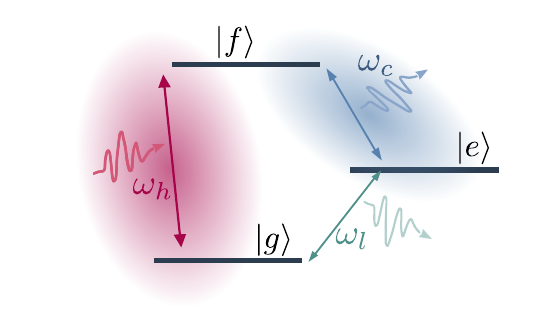}
    \caption{A single three-level system is coupled to two single mode reservoirs with frequencies $\omega_h$ and $\omega_c$. Population inversion is achieved when $\omega_h\beta_h<\omega_c \beta_c$ and in this case an energy current can be emitted through the transition of frequency $\omega_l$ if the system is coupled to a dissipative load or is externally driven.}
    \label{Fig:system_illust}
\end{figure}

Quantum thermal machines offer the prospect of using quantum systems as their working media, with potential performance gains resulting from genuine non-classical properties.
The simplest quantum system which can act as an autonomous heat engine has three energy levels: two transitions exchange energy with a hot and cold heat bath, and the third allows for work, as depicted in Fig.~\ref{Fig:system_illust}.
In one of the earliest contributions to the subject, Scovil and Schulz-DuBois (SSDB) showed how a three-level laser (or maser) can thus be viewed as a heat engine~\cite{scovil_three-level_1959}.
Since then, the thermodynamics of a three-level system have been studied theoretically with Lindblad master equations including situations with an external driving field~\cite{geva_three-level_1994,boukobza_three-level_2007,astafiev_ultimate_2010,kosloff_quantum_2014,singh_2020,kalaee_positivity_2021,Li2017,Gelbwaser2015Power}, and for many copies~\cite{Gelbwaser2019Cooperative}.

A working medium composed of many particles may also show collective effects -- that is, behaviour differing from a set of independent particles.
Superradiance is a notable example in which an ensemble of $n$ dipoles can emit light with power scaling quadratically rather than linearly in $n$~\cite{gross_superradiance_1982}.
This mechanism has recently been proposed to obtain similar enhancements in heat engines and refrigerators~\cite{niedenzu_cooperative_2018,Jaramillo_2016,Macovei2022,Uzdin2016,Myers2020,Myers_2022,Mukherjee_2021,latune_collective_2020,latune_thermodynamics_2019,Souza2021,Kloc2021,Kaimimura2022,Mayo2022}.
Such collective effects may be observed in systems including cold atoms~\cite{Norcia2016Superradiance,Norcia2018Cavity,Kim2022Photonic}, trapped ions~\cite{Casabone2015Enhanced}, and artificial atoms~\cite{Scheibner2007Superradiance}.
Furthermore, one can investigate many-body engines with interactions~\cite{Li_2018,Chen_2019,Keller2020,Fogarty_2020,Carollo2020a,Carollo2020b,Myers2021,Watanabe2020Quantum,Jaramillo_2016,Macovei2022,Li_2018,Halpern2019,Chen_2019,Keller2020,Carollo2020a,Carollo2020b,Myers2021,Williamson2024}, including systems operating around phase transitions~\cite{Fusco2016,Campisi2016,Fogarty_2020,Piccitto2022}.

An important aspect of collective effects is the indistinguishability of the particles in the ensemble.
Physically, this corresponds to the equivalence of the particles as seen by external control fields and environments, expressed mathematically as a symmetry of the dynamics under particle permutations.
Exchange symmetry and anti-symmetry are found in fundamentally identical particles and give rise to Bose-Einstein and Fermi-Dirac statistics, with profound consequences for the thermodynamics of quantum matter.
However, more general exchange symmetry behaviour can emerge in ensembles of indistinguishable particles~\cite{Messiah1964Symmetrization,Tichy2017Extending,yadin_thermodynamics_2023}.
These have implications for many-particle interference~\cite{Brunner2023Many}, quantum Gibbs mixing~\cite{Yadin2021Mixing}, synchronisation~\cite{Solanki2024Exotic}, and non-abelian symmetries associated with non-commuting conserved quantities~\cite{Majidy2023Noncommuting,Kranzl2023Experimental}.

Here, we investigate thermodynamics in collective ensembles of three-level particles as a generalisation of the single-particle SSDB model.
In contrast to a model in which every particle interacts independently with the heat baths, in a collective system, each absorption or emission process is a superposition of all single-particle events (the open-system model is described in detail in Sec.~\ref{sec:model}).
Recent studies involving this model~\cite{Macovei2022,kolisnyk_performance_2023,kolisnyk2024floquet} have focused primarily on states with bosonic exchange symmetry, since this is most relevant for superradiance and moreover simplifies the analysis.
We rather focus on implications of generalised exchange symmetries, motivated in part by recent results demonstrating that the work output of an Otto engine cycle can be optimised by leveraging symmetries outside of the bosonic subspace~\cite{yadin_thermodynamics_2023}. 
In fact, a lasing medium comprised of particles that thermalise collectively, but are not fundamentally identical, may be in a non-bosonic state, but rather in a mixture of different symmetry classes.

We extend the representation-theoretic tools from Ref.~\cite{yadin_thermodynamics_2023} to provide a systematic framework for describing such systems coupled simultaneously to different heat baths.
This general approach further lets us describe systems prepared in any initial state, offering ways to prove certain useful facts about steady states of the dynamics, as well as a significant reduction in the state space needed for numerical computations.
We also aim to understand which features of the single-particle case extend to the collective case, and which are different.
Our analysis involves looking at different quantifiers of work, including unitary work extraction from steady states, and properties of the emitted field via the input-output formalism.

In Sec.~\ref{sec:SU3 intro}, we first introduce the necessary mathematical tools and define the model. The collective three-level model is introduced in Sec.~\ref{sec:general_open} and the general structure of the steady state is given.
The thermodynamics of the system are then discussed in Sec~\ref{sec:Thermo}, first for the steady-state work value, quantified in terms of unitary work extraction of a system only coupled to heat baths. Here, we discuss low-temperature limits and scaling with particle number.
Next, for the energy emitted by an undriven heat engine coupled to a 1-D waveguide as a dissipative load, and finally for a system undergoing periodic weak driving, we show that although collective efficiency is the same as for individual particles, different exchange symmetries exhibit optimal performance according to other figures of merit, for different parameter ranges.  We conclude in Sec.~\ref{sec:conclusions}.

\section{Mathematical preliminaries}\label{sec:SU3 intro}
The dynamics of a single $d$-level system are governed by the special unitary group $\mathrm{SU}(d)$, describing all unitary transformations (up to a global phase).
For an ensemble of $n$ such systems (particles), a second group comes into play: the symmetric group $\mathrm{S}_n$, which describes permutations of the particles.
We aim to understand the behaviour of ensembles undergoing permutation-invariant dynamics, meaning dynamics generated by interactions that look the same under any relabelling of particles.
Physically, this corresponds to ensembles of particles that are indistinguishable to any control fields and environments.
An efficient analysis of these systems is enabled by representation-theoretic structures involving the two aforementioned groups~\cite{yadin_thermodynamics_2023}.
This generalises previous approaches involving spin ensembles~\cite{niedenzu_cooperative_2018,latune_thermodynamics_2019,kolisnyk_performance_2023,kolisnyk2024floquet}.
While these structures apply generally to $\mathrm{SU}(d)$, here we focus on $\mathrm{SU}(3)$ for three-level particles, representing the simplest case in which simultaneous interactions with different environments can be included.
We first introduce the relevant mathematical tools below.

\subsection{Schur Basis}\label{sec:schur_basis}

A representation of $\mathrm{SU}(3)$ associates every $u \in \mathrm{SU}(3)$ with a unitary matrix $U(u)$ such that $U(u)U(u')=U(uu')$, respecting the group structure. The Hilbert space of $n$ three-level systems can be written as $\mathcal{H}^n=(\mathbbm{C}^3)^{\otimes n}$, with the corresponding product representation $U(u) = u^{\otimes n}$ -- i.e., non-interacting collective unitary rotations. This representation is, in general, reducible. That is, there exist non-trivial subspaces in which the action of all $U(u)$ leaves the subspace invariant. There exists a basis of the Hilbert space such that the unitaries can be brought into the block diagonal form   
\begin{equation} \label{eq:unitary_irreps}
    U(u)=\bigoplus_{\lambda} \bigoplus_{i=1}^{m_\lambda} U^{\lambda}(u),
\end{equation}
such that $U(u)$ is a direct sum of irreducible (not reducible) representations and each irreducible representation (irrep) is labelled by an index $\lambda$~\cite{greiner_quantum_1994,goodman_symmetry_2009,zee_group_2016}, counted with multiplicity $m_\lambda$.

Although the form of the decomposition in Eq.~\eqref{eq:unitary_irreps} is generic, \emph{Schur-Weyl duality}~\cite{harrow_applications_2005} gives additional structure due to the product representation of $\mathrm{SU}(3)$.
This duality holds between the actions of the special unitary group and the permutation group $\mathrm{S}_n$ which exchanges particles.
The latter representation $P(\sigma), \, \sigma \in \mathrm{S}_n$, commutes with the former: $[U(u), P(\sigma)] = 0 \; \forall u,\sigma$, implying that there must be a basis simultaneously decomposing both group actions into their irreps.
In fact, there is a basis such that
    \begin{align}\label{eq:block_diag}
        \mathcal{H}^n & =\bigoplus_{\lambda}\mathcal{K}^{\lambda}\otimes \mathcal{H}^{\lambda}, \nonumber \\
        U(u) & = \bigoplus_\lambda \mathbbm{1}_{\mathcal{K}^\lambda} \otimes U^\lambda(u) \, \nonumber \\
        P(\sigma) & = \bigoplus_\lambda P^\lambda(\sigma) \otimes \mathbbm{1}_{\mathcal{H}^\lambda}.
    \end{align}
States in this Hilbert space can be decomposed into the \textit{Schur basis}~\cite{harrow_applications_2005,Harrow2006} labelled by $\ket{\lambda,s_{\lambda},o_{\lambda}}$ such that operators acting on irreps of $\mathrm{S}_n$ change only $s_{\lambda}$ and operators acting on the irreps of $\mathrm{SU}(3)$ change only $o_{\lambda}$. 
Each $\lambda=[\lambda_1,\lambda_2,\lambda_3]$ is associated with one of the partitions of $n = \lambda_1 + \lambda_2 + \lambda_3$ into three non-negative integers.

As noted above, we consider permutation-invariant systems, which means restricting our attention to observables $A$ that remain unchanged under particle relabelling, i.e., $[A, P(\sigma)] = 0 \; \forall \sigma \in \mathrm{S}_n$.
Due to Schur's lemma, this is equivalent to considering all observables of the form $A = \bigoplus_\lambda \mathbbm{1}_{\mathcal{K}^\lambda} \otimes A^\lambda$ in the Schur basis.
One sees that the permutation factors $\mathcal{K}^\lambda$ can be effectively removed from our description of the system; tracing out the degrees of freedom $s_\lambda$ from the Schur basis gives a reduced basis $\ket{\lambda, o_\lambda}$ for an \emph{accessible Hilbert space}
\begin{equation} \label{eq:accessible_hilbert}
    \mathcal{H}^\lambda_\text{acc} = \bigoplus_\lambda \mathcal{H}^\lambda.
\end{equation}
So, ultimately, the relevant state space is described by irreps of $\mathrm{SU}(3)$ without multiplicity.

We end this section with an example to illustrate how the decomposition works for $n=4$ particles. 
A detailed description of the general procedure is given in App.~\ref{app:young_symmetrizer}.
The Hilbert space is broken down into four irreps ${\lambda_a=[4,0,0], \lambda_b=[3,1,0], \lambda_c=[2,2,0], \lambda_d=[2,1,1]}$ -- corresponding to the four allowed permutation symmetry classes for $n=4$. These partitions can be illustrated graphically using Young diagrams, formed from $n$ boxes with rows of lengths given by the parts of $\lambda$ (from top to bottom).
This gives the following four Young diagrams:
\begin{align}
\lambda_a=&\,\ydiagram{4}
& \lambda_b=&\,\ydiagram{3,1} \\\nonumber
\lambda_c=&\, \ydiagram{2,2}
& \lambda_d=&\,\ydiagram{2,1,1}.
\end{align}
The number of ways each of these diagrams can be filled with the numbers \{1,2,3,4\} without repetition and with numbers strictly increasing along both rows and columns gives the multiplicity $m_{\lambda}$ of each $\mathrm{SU}(3)$ irrep, or equivalently, the dimension of the corresponding $S_4$ irrep. The number of ways each of these diagrams can be filled with the numbers $\{0,1,2\}$, now with repetitions allowed, with numbers strictly increasing along columns and not decreasing along rows, gives the dimension $d_\lambda$ of the $\mathrm{SU}(3)$ irrep.
The filling prescription is consistent with the fact that there are no more than three rows.
Additionally, each of these fillings represents a state that has a particular type of permutation symmetry. The states constructed in this way are linearly independent and states with different Young diagrams are orthogonal.
The symmetry type of the state can be read from the diagram. The diagram prescribes how states from the product basis consisting of the single-particle states $\ket{0},\ket{1},\ket{2}$ are combined either symmetrically (along rows), antisymmetrically (along columns) or a combination of both (for diagrams with both rows and columns) to construct the reduced Schur basis of a particular irrep~\cite{Harrow2006}. The details of this are given in App.~\ref{app:young_symmetrizer}.
For example, the basis of $\lambda_a$ describes all fully symmetric states since its Young diagram contains only a row, while the states of $\lambda_b$ $\lambda_c$ and $\lambda_d$ have a mixture of anti-symmetry and symmetry.
The subspace defined by $\lambda_d$ admits $d_{\lambda_d}=3$ fillings and is equivalent to the Hilbert space of a single particle, which is represented by a Young diagram with a single box. From this, we see that all columns with three boxes can be dropped from the description. 

Therefore, in general we label the irreps using the `highest weight' notation: We replace $\lambda$ with $(p,q)$, which count the number of columns of length 1 and 2 respectively in the Young diagram. Using this notation we can also explicitly give the dimension and multiplicity of a given irrep of a system of $n$ particles~\cite{yadin_thermodynamics_2023,greiner_quantum_1994},
\begin{align} \label{eq:irrep_dims}
    d_\lambda=&\frac{1}{2}(p+1)(q+p+2)(q+1), \\\nonumber
    m_\lambda=&\frac{2d_\lambda n!}{\bigg(\frac{n+q-p+3}{3}\bigg)!\bigg(\frac{n+q+2p+6}{3}\bigg)!\bigg(\frac{n-2q-p}{3}\bigg)!}.
\end{align}
\subsection{$\mathrm{SU}(3)$ generators and the open system model} \label{sec:model}
The block-diagonal structure of the accessible Hilbert space given in Eq.~\eqref{eq:accessible_hilbert} allows one to speed up calculations of the dynamics of larger systems considerably.
In particular, $\lambda$ is a conserved quantity under permutation-invariant dynamics -- so each irrep block can be treated independently.
Here, we describe the basic dynamical variables needed for this description, which are used later to define our open-system model.
The dynamical operators are generators of $\mathrm{SU}(3)$. Two of them are diagonal and we label them by $W_z$ and $Y$~\cite{greiner_quantum_1994}.
For a single particle they are diagonal in the computational basis $\{\ket{0},\ket{1},\ket{2}\}$:
\begin{equation}
    W^{(1)}_z=\frac{1}{2}\begin{pmatrix}
1 & 0 & 0\\
0 & -1 & 0\\
0 & 0 & 0
\end{pmatrix}, \quad
Y^{(1)} =\frac{1}{3}\begin{pmatrix}
1 & 0 & 0\\
0 & 1 & 0\\
0 & 0 & -2
\end{pmatrix},
\end{equation}
and are proportional to two of the Gell-Mann matrices.
The remaining six non-diagonal generators come in pairs, denoted $W_+=W_-^\dagger$, $U_+ = U_-^\dagger$, and $V_+ = V_-^\dagger$ which act as the $\mathrm{SU}(3)$ equivalent of the more familiar $\mathrm{SU}(2)$ raising and lowering operators; their single-particle versions are 
\begin{align}
    W^{(1)}_+ = \begin{pmatrix}
        0 & 1 & 0 \\
        0 & 0 & 0 \\
        0 & 0 & 0
    \end{pmatrix}, \;
    U^{(1)}_+ = \begin{pmatrix}
        0 & 0 & 0 \\
        0 & 0 & 1 \\
        0 & 0 & 0
    \end{pmatrix}, \;
    V^{(1)}_+ = \begin{pmatrix}
        0 & 0 & 1 \\
        0 & 0 & 0 \\
        0 & 0 & 0
    \end{pmatrix}.
\end{align}
Any other elements of the Lie algebra $\mathrm{su}(3)$ can be obtained from linear combinations of these 8 operators.
All their commutators are listed in App.~\ref{app:SU3}. 
For the $n$-particle product representation, the collective generators are sums of single-particle terms: $A = \sum_{i=1}^n A^{(1)}_i$, where $A^{(1)}_i$ is any one of the above generators acting on particle $i$.

Finally, we introduce a labelling of the reduced Schur basis which transforms conveniently under the generators~\cite{de_swart_octet_1963} and will allow us to simplify the calculations of the steady state behaviour.
This basis $\ket{(p,q),W,w,y}$ is characterised by the irrep label $\lambda = (p,q)$, plus three additional quantum numbers.
\begin{figure}
\centering
   \includegraphics[width=0.35\textwidth]{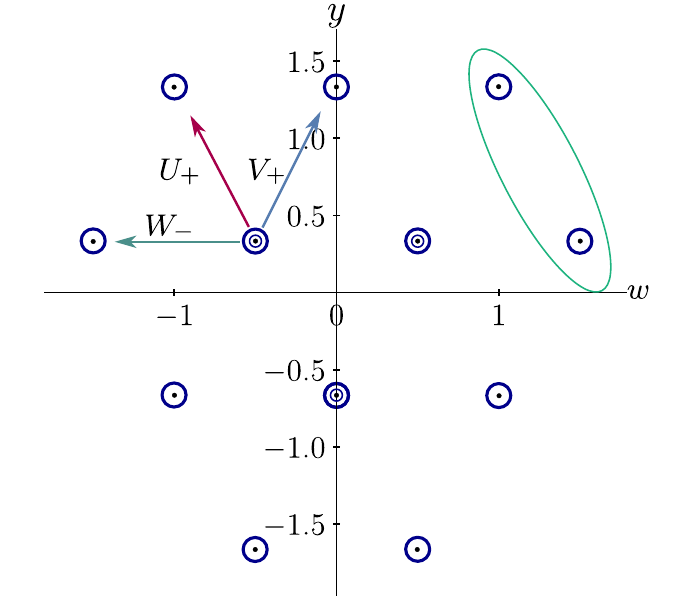}
    \caption{A diagram of the reduced basis states $\ket{(p,q),W_i,w_i,y_i}$ of one of the irreducible representations found for 4 three-level systems, labelled $\lambda = (p,q) =(2,1)$. The values of $w_i$ are on the $x$-axis and $y_i$ on the $y$-axis. The central states are doubly degenerate, which is shown with concentric circles. These states have different values for $W_i$. The actions of the $U_{\pm}$, $V_{\pm}$ and $W_{\pm}$ operators on states are indicated in the diagram in the upper left corner. The steady state achieved when $\omega_c\beta_c\gg1$ is confined to a mixture of the states in the upper right diagonal of any weight diagram, here these states are circled in green.}  \label{Fig:basis}
\end{figure}
The number $W$ labels the eigenstates of the operator $W^2 = \frac{1}{2}(W_+ W_- + W_- W_+) + W_z^2$ -- i.e., the total spin-length quantum number defining the $\mathrm{SU}(2)$ sub-irreps arising from the $W_z, \, W_\pm$ operators.
The remaining quantum numbers $w$ and $y$, known as \emph{weights}, label the eigenstates of the commuting generators $W_z$ and $Y$. The actions of the generators on these states and their effects on $w$ and $y$ are are given in detail in App.~\ref{app:SU3}. This is illustrated graphically in a \emph{weight diagram} as shown in Fig.~\ref{Fig:basis}, where the basis elements of irrep corresponding to $(p,q)=(2,1)$ are plotted with the $x$-coordinate corresponding to the values of $w$ and the $y$-coordinate to the value of $y$.
The central points in any hexagonal diagram are degenerate, forming a \emph{weight space}, which is illustrated by concentric circles in the figure. These points have the same single particle expectation values with a non-interacting Hamiltonian and therefore, have the same $w$ and $y$ but different $W$ quantum numbers.

The irrep diagrams have six (or three) sides containing $(p+1)$ points on the top edge and $(q+1)$ points on the bottom, with a $2\pi/3$ rotational symmetry, up to an axis rescaling.
The right-most point (with largest $w$) is known as the highest-weight vector, and has coordinates $([p+q]/2,[p-q]/3)$.

\section{General open system dynamics} \label{sec:general_open}

In this section, we introduce the general form of the open-system master equations that will be studied in different varieties in Sec.~\ref{sec:Thermo}, and discuss the form of their steady states.
The single-particle three-level maser depicted in Eq.~\eqref{Fig:system_illust} has transitions resonant with two single-mode bosonic reservoirs at a high and a low temperature with positive frequencies $\omega_h$, $\omega_c$ respectively, and a lasing transition $\omega_l = \omega_h - \omega_c$, described by the Hamiltonian
\begin{equation} \label{eq:single_particle_h}
    h=\omega_l\ket{e}\bra{e}+\omega_h\ket{f}\bra{f},
\end{equation}
where we set $\hbar=1$. The energy eigenstates $\{\ket{g},\ket{e},\ket{f}\}$ map onto the computational basis states $\{\ket{1},\ket{0},\ket{2}\}$.
It is possible to generalise the single three-level maser to a collective ensemble of three-level systems using the $\mathrm{SU}(3)$ generators introduced in the previous section. The free Hamiltonian of the ensemble is
\begin{equation}\label{eq:free_hamiltonian}
    H_S=\omega_lW_z-\left(\frac{\omega_c+\omega_h}{2}\right)Y.
\end{equation}
This is equivalent (up to an unimportant constant shift) to the non-interacting Hamiltonian where each particle individually has the Hamiltonian in Eq.~\eqref{eq:single_particle_h}. The Schur basis vectors are eigenvectors corresponding to energy eigenvalues $E_{wy}^\lambda = \omega_l w - (\omega_c+\omega_h) y/2$, which may be degenerate with respect to different $W$ values.
The ensemble is chosen to interact with bosonic thermal baths, with inverse temperatures $\beta_c,\, \beta_h$, via the collective jump operators $V_\pm$ and $U_\pm$, respectively.
For example, $V_+$ describes an exchange of energy $\omega_c$ from the system to the cold bath, occuring as a superposition of all single-particle transitions from $\ket{f}$ to $\ket{e}$. 
Assuming the standard weak coupling, Born-Markov, and secular approximations, these collective dynamics can be modelled using a Lindblad master equation, taking the form in the interaction picture~\cite{yadin_thermodynamics_2023}
\begin{align}\label{eq:full_master}
      &\dot{\rho}=g_u\mathcal{L}_U(\bar{n}_h)\rho+g_v\mathcal{L}_V(\bar{n}_c)\rho+g_w\mathcal{L}_W(\bar{n}_l)\rho
\end{align}
where,
\begin{equation}
    \mathcal{L}_O(\bar{n}_i)\rho= \bar{n}_i\mathcal{D}[{O_{-}}]\rho+g_i(\bar{n}_{i}+1)\mathcal{D}[{O_{+}}]\rho.
\end{equation}
Here,
$\mathcal{D}[{O}]\rho={O}\rho {O}^{\dagger} -\frac{1}{2}\{{O}^{\dagger}{O},\rho\} $ and $\bar{n}_{i}=1/(\textrm{exp}(\omega_{i}\beta_{i})-1)$, is the bosonic average occupation of the mode with frequency $\omega_{i}$ of the reservoir with inverse temperature $\beta_{i}=1/k_BT_{i}$ .
 
While Eq.~\eqref{eq:full_master} applies to the state of the entire system, we can similarly describe the dynamics of each irrep block.
In the accessible state space, we generally write $\rho_\text{acc} = \bigoplus_\lambda p^\lambda \rho^\lambda$, where $p^\lambda$ is a probability distribution over the index $\lambda$ and $\rho_\lambda$ is a normalised state on $\mathcal{H}^\lambda$.
(A generic state may have coherences between blocks when reduced down to the accessible Hilbert space, but these are irrelevant and can be ignored under permutation-invariant dynamics~\cite{yadin_thermodynamics_2023} since they are not detectable by permutation invariant observables. Furthermore, in the steady state these coherences decay to zero -- see App.~\ref{app:steady_state}.) 
As noted above, since $\lambda$ is conserved, the irreps do not mix, so we can write down a master equation for each $\rho^\lambda$ of exactly the same form as Eq.~\eqref{eq:full_master} -- where each generator $A \in \{W_\pm,U_\pm, V_\pm\}$ is replaced by its component $A^\lambda$ on the irrep, which can be calculated by projecting onto the reduced basis discussed in Sec.~\ref{sec:schur_basis} and App.~\ref{app:young_symmetrizer}. 
For simplicity of notation, we will often not explicitly write down the label $\lambda$, always implicitly working within a fixed irrep.

We are mainly interested in properties of the steady states under the master equations of the form of Eq.~\eqref{eq:full_master}.
The probabilities $p^\lambda$ are set by the initial state and remain fixed in time; they can in principle be chosen arbitrarily with an appropriate state preparation.
For example, one possible choice of preparation is to fully thermalise each particle independently with respect to some inverse temperature $\beta_0$: $\rho_{\beta_0} = \exp(-\beta_0 H_S)/Z_{\beta_0}$, where $Z_{\beta_0} = \tr[ \exp(-\beta_0 H_S)] = \tr[\exp(-\beta_0 h)]^n$.
In the reduced Schur basis, this state becomes
\begin{align} \label{eq:initial_thermal}
    \rho_{\beta_0, \text{acc}} = \bigoplus_\lambda p^\lambda \, \frac{e^{-\beta_0 H_S^\lambda}}{Z^\lambda_{\beta_0}}, \quad p^\lambda = \frac{m_\lambda Z^\lambda_{\beta_0}}{Z_{\beta_0}}.
\end{align}
For example, in the limit $\beta_0 \to \infty$, the system approaches the ground state, which is contained in the fully symmetric subspace labelled by $\lambda = [n,0,0]$, equivalently $(p,q) = (n,0)$.
In the high-temperature limit $\beta_0 \to 0$, $Z^\lambda_{\beta_0} \to d_\lambda$, so $p^\lambda \to m_\lambda d_\lambda / 3^n$.

Within each irrep, there is always a unique steady state $\rho^\lambda_\infty$, except for the degenerate cases where two or more of the $g_i$ coefficients vanish~\cite{yadin_thermodynamics_2023}.
In App.~\ref{app:steady_state}, we show the following:
\begin{res} \label{res:steady_diagonal}
    The steady state $\rho^\lambda_\infty$ has vanishing coherences between different weight spaces corresponding to the quantum numbers $(w,y)$:
    \begin{align}
        \mel{\lambda,W,w,y}{\rho^\lambda_\infty}{\lambda,W',w',y'} = 0 \;\text{ if } w \neq w' \text{ or } y \neq y'.
    \end{align}
\end{res}
This result significantly simplifies calculations involving the steady state, requiring us to keep track of coherences only between differing $W$ values.

Since the system's energy eigenvalues are functions of $w$ and $y$, this implies vanishing coherence between different energies.
However, Result~\ref{res:steady_diagonal} is a stronger statement in the case that the Hamiltonian has degeneracies between different weight spaces (which can happen if the energy gaps are rationally dependent).

\section{Thermodynamics in three models}\label{sec:Thermo}

In this section, we assess the performance of a collective three-level engine based on the work output characteristics in three different models.

\subsection{Unitary work extraction from steady state}

\begin{figure}
\centering
   \includegraphics[width=0.35\textwidth]{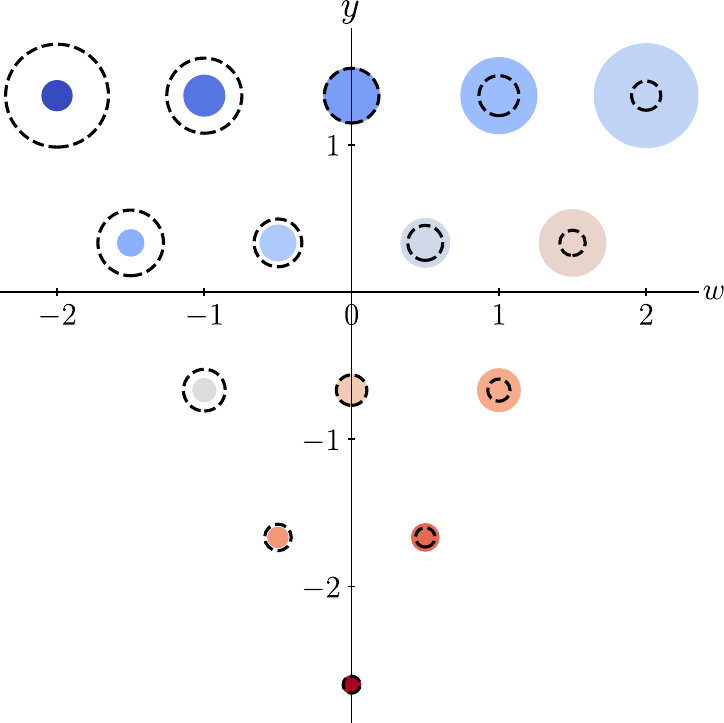}
    \caption{Plots of the populations of the steady state of Eq.~\eqref{eq:ergotropy_master} for the irrep $(p,q)=(4,0)$ are shown here with shaded circles. The hot and cold reservoirs have frequencies and inverse temperatures $\omega_c=2/3$, $\omega_h=5/3$  $\beta_c=1.5$ and $\beta_h=0.8$. The size of the circle at point $(w_i,y_i)$ is proportional to the population of the state $\ket{(4,0),W_i,w_i,y_i}$. The state $\tilde{\rho}_{W_x}$ formed by rotating under $W_x$ is shown with dashed black circles. The shading of the circles indicates the magnitude of the eigenvalue of the Hamiltonian at each point, increasing from dark blue to dark red. All system parameters here and in following figures are in units of $\omega_l$.} \label{Fig:population_plot}
\end{figure} 
\begin{figure}
\centering
   \includegraphics[width=0.45\textwidth]{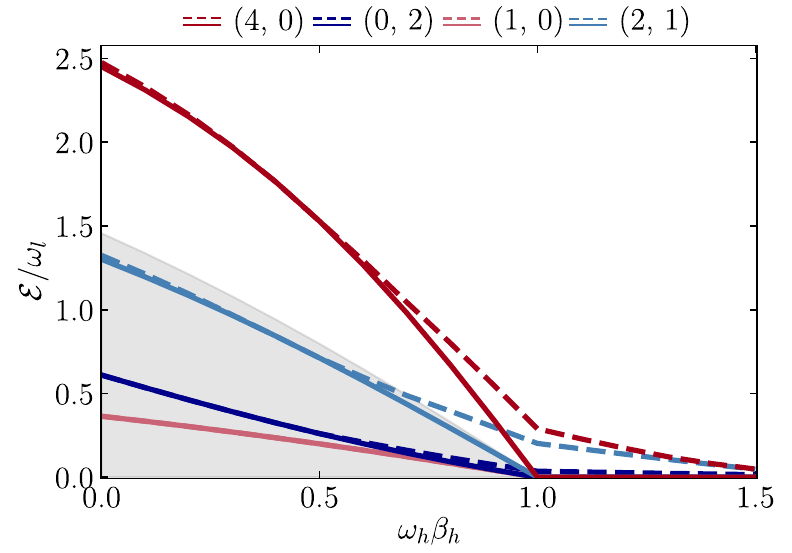}
    \caption{Energy extracted from a collective system of 4 particles within various irreps of $\mathrm{SU}(3)$ denoted $(p,q)$ in the key above the plot. The lasing ergotropy $\mathcal{E}_l$ (solid lines) is compared to the total energy that could be extracted from a unitary that constructs a completely passive state (dashed lines). The upper border of the shaded region shows $\mathcal{E}_l$ for 4 independently coupled 3-level systems, thus when irreps fall within the shaded region they perform worse than independently coupled systems.  Here they are fixed to $\omega_c=2/3$, $\omega_h=5/3$,  $\beta_c=1.5$,  $g_u,g_v=0.1$.}  \label{Fig:extracted_energy_n_4}
\end{figure}  

First, we study the energy that could be extracted from a system that has reached its steady state under the dynamics of Eq.~\eqref{eq:ergotropy_master} with the coupling to a hot bath and cold bath and we assume no dynamics on the $\omega_l$ transition. That is, 
\begin{align}\label{eq:ergotropy_master}
    \dot{\rho}=g_u\mathcal{L}_U(\bar{n}_h)\rho+g_v\mathcal{L}_V(\bar{n}_c)\rho.
\end{align}
This is motivated in parallel to the single-particle SSDB master equation,
\begin{align}
\dot{\rho}_{\text{sing}}=&g_v(\bar{n}_c\mathcal{D}[{\ketbra{e}{f}}]\rho+(\bar{n}_{c}+1)\mathcal{D}[{\ketbra{f}{e}}]\rho \nonumber) \nonumber\\
      &+g_u(\bar{n}_h\mathcal{D}[{\ketbra{g}{f}}]\rho + (\bar{n}_h+1)\mathcal{D}[{\ketbra{f}{g}}]\rho),
\end{align}
which gives a thermodynamical interpretation of the lasing threshold condition via population inversion between the levels $|g\ra$ and $|e\ra$.
In a given irrep for the collective case, the total amount of population inversion among the available energy levels is most straightforwardly characterised by the ergotropy 
\cite{allahverdyan_maximal_2004}, defined as the maximum energy that can be extracted under unitary operations:
\begin{align}
    \mathcal{E}(\rho) := \max_U \, \left( \tr[H_S \rho] - \tr[H_S U \rho U^\dagger] \right).
\end{align}
The optimal unitary achieving the ergotropy maps the state $\rho = \sum_k r_k \dyad{\psi_k}$ onto its corresponding \emph{passive state} $U\rho U^\dagger = \tilde\rho := \sum_k r_k^\downarrow \dyad{E_k}$, which is diagonal in the energy eigenbasis and has eigenvalues decreasing with respect to increasing energy. 
For energy-diagonal $\rho$, the ergotropy can be understood as a measure of population inversion as $U$ rearranges the populations such that the $r_k^{\downarrow}$ are in descending order with respect to ascending $E_k$-values.
Note that $U$ is not constrained here to the $\mathrm{SU}(3)$ product representation; it must only be permutation-invariant, so is completely general within each irrep $\lambda$.

In practice, the optimal unitary $U$ may not be achievable if energy extraction is limited to merely the lasing transition.
The operator providing rotations in this subspace is $W_x = W_+ + W_-$, so we define the more restrictive \emph{lasing ergotropy} 
\begin{align}
    \mathcal{E}_l(\rho) := \max_\theta \, \left( \tr[H_S \rho] - \tr[H_S e^{-i\theta W_x} \rho e^{i\theta W_x}] \right). \label{eq:lasingErgotropy}
\end{align}

We first give two facts about the ergotropy of the steady state:
\begin{res} \label{res:ergotropy}
    \begin{enumerate}
        \item[(i)] $\mathcal{E}_l(\rho^\lambda_\infty) = 2 \omega_l \max \{ \tr[W_z \rho^\lambda_\infty], 0 \}$, and when it is non-zero, the optimal rotation angle is $\theta = \pi$.

         \item[(ii)] In the limit of a very low-temperature cold bath, $\omega_c \beta_c \gg 1$, the steady state $\rho^\lambda_\infty$ is constrained to the upper right-hand diagonal including $q+1$ states circled in green in Fig.~\ref{Fig:basis}.
         In this case, $\mathcal{E}_l(\rho^\lambda_\infty)=\mathcal{E}(\rho^\lambda_\infty)$ if $q \leq \omega_l / \omega_c$. 
    \end{enumerate}
  
\end{res}
Part (i) can be shown as follows: under the rotation generated by $W_x$, the operator $W_z$ transforms in the Heisenberg picture as $W_z \to \cos\theta W_z + \sin\theta \, W_y$.
Then, since $Y$ is unchanged by this rotation, the lasing ergotropy \eqref{eq:lasingErgotropy} simplifies to  
\begin{align}
    \mathcal{E}_l(\rho_\infty^\lambda) = \omega_l \max_\theta \, \tr \left\{ \left[ (1-\cos\theta)W_z - \sin\theta W_y \right] \rho  \right\} .
\end{align}

Moreover, since $\rho$ is diagonal in $w$ by virtue of Result~\ref{res:steady_diagonal}, it is invariant under rotations generated by $W_z$, and so $\ev{W_y} := \tr [W_y\rho] = 0$.
Then either $\ev{W_z} \leq 0$, in which case no energy is extractable, or $\ev{W_z} > 0$ in which case it is best to rotate $W_z \to -W_z$. 
(Notice that, even though we restricted lasing ergotropy to rotations generated by $W_x$ for simplicity, the result (i) also holds for more general rotations generated by any combination of $W_x, W_y, W_z$.)

Part (ii) is proven in App.~\ref{app:cold}.
In this low-temperature limit, the state has a thermal distribution over the upper-right line in Fig.~\ref{Fig:basis}, behaving effectively as a spin-$q/2$ system at the temperature of the hot bath. The lasing ergotropy then also coincides with the full ergotropy, and is found to be
\begin{align} \label{eq:cold_ergotropy}
    \mathcal{E}_l(\rho_\infty^\lambda) & = \omega_l \left[ p + q + \frac{1}{1-e^{-\beta_h\omega_h}} - \frac{q+1}{1 - e^{-(q+1)\beta_h\omega_h}} \right] \nonumber \\
    & ( \text{for } \omega_c \beta_c \gg 1).
\end{align}

For general temperatures, it is difficult to give a full description of the steady state -- in contrast to the single-particle case and the fully symmetric subspace~\cite{Macovei2022}.
However, we observe numerically that the single-particle population inversion condition, $\omega_c \beta_c > \omega_h \beta_h$, is also the condition  for nonzero lasing ergotropy in every irrep.

The populations of an exemplary steady state $\rho$ of the $(4,0)$ irrep and the corresponding lasing-passive state $\tilde{\rho}_{W_x}=e^{-i\theta W_x} \rho e^{i\theta W_x}$ are illustrated in Fig.~\ref{Fig:population_plot} by the shaded circles and dashed black circles respectively.
Obtaining a completely passive state from the steady state $\rho$ would require transitions diagonally in the diagram as well as horizontally, whilst obtaining the state $\tilde{\rho}_{W_{x}}$ from $\rho$ requires horizontal transitions corresponding to operations under $W_{\pm}$ only. 
Indeed, there are situations where the state $\tilde{\rho}_{W_{x}}$ will have a smaller energy difference from the steady state than the completely passive state does. That is, one could extract more work from the steady state than just from the lasing transition alone, and this additional work resource can extend beyond the temperature window of lasing.

This is exemplified for an ensemble of 4 particles in Fig.~\ref{Fig:extracted_energy_n_4}. Given that one prepares a state with a particular initial symmetry, the dynamics are confined to that subspace. Therefore, we show the energy that could be extracted from each subspace. The maximum energy will be extracted from the subspace with the highest value of $p$, and this will always be the fully symmetric subspace, i.e., $p=n$, $q=0$. We also show the energy that would be extracted from $n$ independent particles (top line of the shaded region). 
When $\omega_h\beta_h \ll \omega_c\beta_c$, this is identical to the energy extracted from the fully symmetric subspace. However, when the temperature difference is less extreme, as is shown in Fig.~\ref{Fig:extracted_energy_n_4}, independent particles do not produce as much energy as collectively coupled ones do.
\begin{figure}
\centering
   \includegraphics[width=0.45\textwidth]{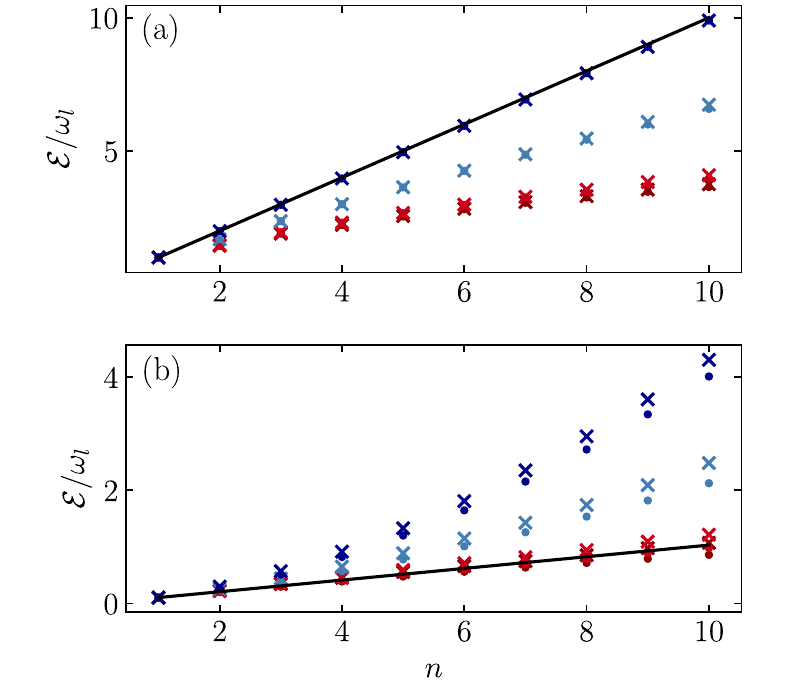}
    \caption{The scaling of the ergotropy (crosses) and lasing ergotropy (points), for an ensemble prepared in an initially thermal state $\rho_{\beta_0,\text{acc}}$ is compared here to the ergotropy of $n$ individual particles (black line). The temperature of the initial state is indicated by the colour of the markers with $\beta_0=\{0.09,0.45,1.5,5.0\}$ going from dark red to dark blue. In (a), we explore the limit where $\beta_c\to \infty$ and find that it is only possible to extract the same amount of energy as from individual particles when the state is prepared at a sufficiently low temperature. In (b), we show that for intermediate cold bath temperatures ($\beta_c=1.5$) it is possible to perform better than individual systems even for moderate $\beta_0$. This effect is enhanced for large $n$. Here, we fix the temperature of the hot bath in both scenarios to $\beta_h=0.45$ and the frequencies of the cold and hot baths are  $\omega_c=2/3$ and $\omega_h=5/3$ respectively and $g_u,g_v=0.1$.  }  \label{Fig:ergotropy_scaling}
\end{figure} 

We can further make statements about the many-particle limit, considering a case where the initial state is prepared as a product of the form $\rho_1^{\otimes n}$.
In the limit of large $n$, this distribution concentrates on subspaces whose Young diagrams have row lengths in proportion with the decreasingly ordered eigenvalues $r_i$ of $\rho_1$: i.e., such that $\lambda_i \approx n r_i$~\cite{Alicki1988Symmetry,Keyl2001Estimating}.
(Related results have been applied in a thermodynamical context in Refs.~\cite{Alicki2013Entanglement,yadin_thermodynamics_2023,Cavina2024Symmetry}.)
This means that we need analyse only a single `typical' irrep with parameters $(p,q) = (n[r_1-r_2], n[r_2-r_3])$; here, we are ignoring fluctuations in $p$ and $q$ of order $\sqrt{n}$.
This greatly simplifies calculations via an exponential reduction of the effective Hilbert space down to a dimension of order $n^3$, as seen in Eq.~\eqref{eq:irrep_dims}.

Extending Result~\ref{res:ergotropy} (ii) to the many-particle limit, we find the lasing ergotropy of the typical irrep in the case $\omega_c \beta_c \gg 1$ by substituting the typical $(p,q)$ values into Eq.~\eqref{eq:lasingErgotropy}, resulting in $\mathcal{E}_l \approx \omega_l (p + 2q) = \omega_l n(1 - 3r_3)$ to leading order in $n$.

In Fig.~\ref{Fig:ergotropy_scaling}, we compare the collective and independent models by choosing an initial state such that $\rho_1$ is the thermal state of the single-particle Hamiltonian at some inverse temperature $\beta_0$, imagining that the system has been prepared by fully thermalising.
The parameter $\beta_0$ can then be used to tune this state via its probability distribution $p^\lambda$ over irreps, as seen in Eq.~\eqref{eq:initial_thermal}.
(Note, however, that this parameter does not affect the case of independent particles.)
For very low cold bath temperatures, Fig.~\ref{Fig:ergotropy_scaling} (a) indicates that the independent case has optimal ergotropy, which is only approached in the collective model by taking large $\beta_0$ -- i.e., preparing in the fully symmetric subspace.
On the other hand, for the intermediate bath temperatures in Fig.~\ref{Fig:ergotropy_scaling} (b), large $\beta_0$ preparations outperform independent particles.
Moreover, this advantage persists even for moderate $\beta_0$, in which irreps outside of the fully symmetric one contribute.
It is noteworthy that the scaling in $n$ shown here is initially super-linear.

Another interesting case to consider is when $\rho_1$ is the steady state of the single-particle model.
If we choose temperatures such that $\omega_c \beta_c = \omega_h \beta_h$, then the populations of levels $\ket{g}$ and $\ket{e}$ are equal -- this single-particle state is just at the lasing threshold, has zero lasing ergotropy, and is passive.
Since $\rho_1$ is not a thermal state, the many-copy state $\rho_1^{\otimes n}$ is non-passive for sufficiently large $n$~\cite{Lenard1978Thermodynamical}, and moreover has positive lasing ergotropy in the collective model.
The many-particle typical irreps are then characterised by $p \approx 0$ and $q \approx n(1 - e^{-\beta_c \omega_c})/(2 + e^{-\beta_c \omega_c})$.
The corresponding weight diagram is a triangle, reflected vertically compared with the fully symmetric irrep.
The collective model then extracts work from a subspace with symmetry type that is in some sense far from bosonic.
This result may be compared with Ref.~\cite{Gelbwaser2019Cooperative}, which also considers work extraction via multi-copy operations, though not collective in the sense of obeying permutation symmetry.

\subsection{Power emission into dissipative load}

The maximum energy that could be extracted from the ensemble (ergotropy) is useful in determining if there is a population inversion at steady state to begin with. However, in analogy to the SSDB heat engine, we want to evaluate the actual power output and efficiency of the system under load. 
For a simple model, consider that, in addition to the two thermal baths of finite temperature with inverse temperatures $\beta_h$ and $\beta_c$, a zero-temperature bath is coupled to the $W_{\pm}$ transition. This bath functions as a dissipative load~\cite{niedenzu_quantum_2018,niedenzu_concepts_2019}. Thus, the system evolution is
\begin{equation}
    \dot{\rho}=g_u\mathcal{L}_U(\bar{n}_h)\rho+g_v\mathcal{L}_V(\bar{n}_c)\rho+g_w\mathcal{D}[{W_-}]\rho.
\end{equation}
 The heat current into the system from the hot reservoir is
  \begin{equation}
    \mathcal{I}_{h}=\tr[g_{u}\mathcal{L}_{U}(\bar{n}_h)\rho H_S]
\end{equation}
with a similar expression for the heat current flowing in from the cold reservoir.
This simplifies to
 \begin{equation} \label{eq:heaw_hot}
\mathcal{I}_{h}=\omega_{h}g_{u}\tr\{[(\bar{n}_{h}+1)U_-U_+-\bar{n}_{h}U_+U_-]\rho\}.
 \end{equation}
with an analogous expression for the cold reservoir with the respective cold bath parameters and operators.
The energy current flowing into the dissipative load~\cite{Mari_2015},
\begin{equation}\label{eq:dissip_power}
    \mathcal{P}=-\tr\{{H_S}g_w\mathcal{D}[{W_-}](\rho)\},
\end{equation}
can be identified as the power.
This expression simplifies in a similar way to the heat current so that
\begin{equation} \label{eq:dissip_power2}
    \mathcal{P}=\omega_l g_{w}\tr\{W_+W_-\rho\}.
\end{equation}
 The ratio $\eta = \mathcal{P}/\mathcal{I}_h$ of these two quantities is then the efficiency of the engine.
We find that $\eta$ depends only on the frequencies of the hot and cold transitions:
\begin{res}
    The steady-state efficiency in every irrep is $\eta = 1 - \omega_c / \omega_h$.
\end{res}
This result was already known for a single three-level system~\cite{singh_2020, kosloff_quantum_2014} but it is remarkable that it extends to every irrep of a collective system.
To see this, consider the operator counting the total number of particles in the ground state $\ket{g}$, which can be written $N_g = n/3 + Y/2 - W_z$.
Then the time-derivative
\begin{align}
    \partial_t \ev{N_g} & = \ev{g_u[(\bar{n}_h+1)U_-U_+-\bar{n}_h U_+U_-]+g_w W_+W_-} \nonumber \\
        & = \frac{\mathcal{I}_h}{\omega_h} - \frac{\mathcal{P}}{\omega_l}
\end{align}
must vanish in the steady state, having used Eqs.~\eqref{eq:heaw_hot} and \eqref{eq:dissip_power2}.
So $\mathcal{P} / \mathcal{I}_h = \omega_l / \omega_h = 1 - \omega_c / \omega_h$.

 \begin{figure}
\centering
   \includegraphics[width=0.45\textwidth]{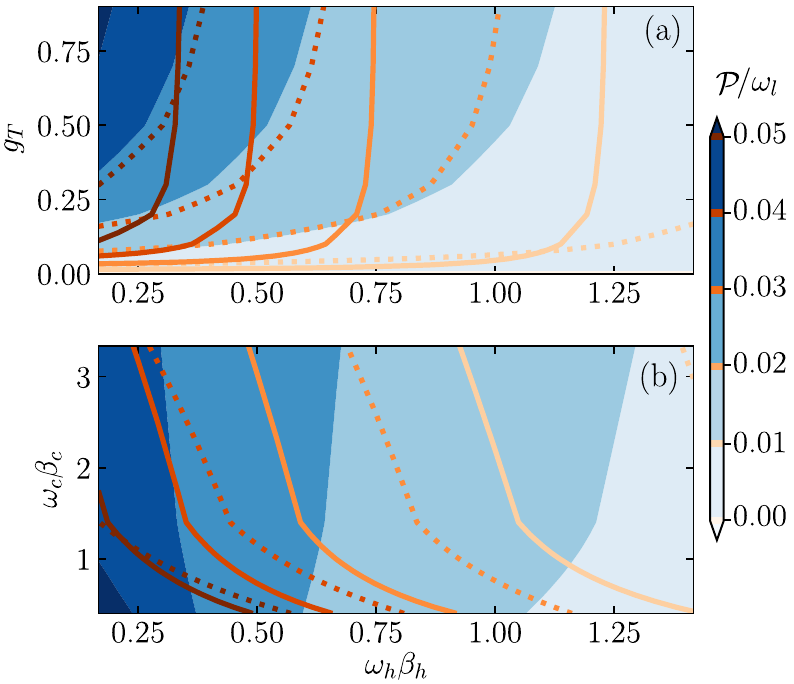}
    \caption{The energy current emitted by the ensemble of three-level system into a zero-temperature bath through the $W_{\pm}$ transition is highest for the fully symmetric subspace and, as shown in (a), for stronger coupling $g_T$ to the dissipative load. The blue shaded regions denote the energy current from a system initialised in the $(2,1)$ irrep and the solid orange contours are for the $(4,0)$ irrep, where the lines in the colour-bar denote the contour of all power values greater than or equal to the assigned colour. The dotted contours denote the energy current of 4 distinguishable particles. Here, the temperature of the cold bath is fixed to $\beta_c=1.1$. The frequencies of the hot and cold baths are set to, $\omega_h=5/3$ and $\omega_c=2/3$. The maximum energy is emitted when the temperature of the cold bath is higher for both irreps, as shown in (b), where the coupling to the load is fixed at $g_T=0.9$ corresponding to the highest emission in (a). Couplings to the hot and cold reservoirs are both set to 0.1. }  \label{Fig:sponw_emission_bh}
\end{figure}
 The energy emitted into the zero-temperature reservoir is depicted for an collection of 4 particles in Fig.~\ref{Fig:sponw_emission_bh}. Here, we show that the fully symmetric irrep (solid orange contours) outperforms the other irreps (here the (2,1) irrep is shown by shaded blue regions). However, it is possible that independent particles may emit more energy than other irreps, as is shown in this figure for four independent particles (dotted lines).    

From Eq.~\eqref{eq:dissip_power}, we see that the energy current emitted by the system is proportional to the coupling strength to the zero temperature bath. Indeed, for a stronger coupling to the dissipative load there appears to be improved performance of the engine. However, the aim is for the system to perform \emph{work} on the load. Therefore, it is not enough to consider only the magnitude of the energy current but also the `quality' of the emitted light -- i.e., to what extent it can be considered work-like or heat-like.

To quantify this, consider coupling the output mode of the system to a 1-D waveguide~\cite{monsel_energetic_2020}. We identify the work-like component of the emitted with the \emph{coherent} part of the output spectrum.
The (normalised) intensity spectrum is~\cite{PhysRevA.40.5516}
\begin{align}
    S(\omega)=&\frac{1}{2\pi}\bigg[\int_{0}^{\infty}d\tau \, G^{(1)}(\tau,\tau)\bigg]^{-1}\\\nonumber
    &\int_{0}^{\infty}d\tau \int_{0}^{\infty}d\tau' \, e^{-i\omega (\tau-\tau')}G^{(1)}(\tau,\tau'),
 \end{align}
where $G^{(1)}(t,t') = \ev{a_\text{out}(t)^\dagger a_\text{out}(t')}$ (using Heisenberg-picture operators) is the first-order Glauber coherence function for the output mode with annihilation operator $a_\text{out}$.
At steady state, this expression simplifies to
\begin{align} \label{eq:power_fraction}
    S(\omega) & = \frac{ \frac{1}{2\pi} \int_{-\infty}^\infty d\tau \, e^{-i \omega \tau} G^{(1)}(\tau)}{G^{(1)}(0)} \nonumber \\
        & =: \frac{P(\omega)}{P_\text{tot}},
\end{align}
where $G^{(1)}(\tau) := G^{(1)}(t+\tau,t)$ for any $t$.
Here, $P(\omega)$ is the photon flux per unit frequency into mode $\omega$ of the waveguide (named the power spectrum in quantum optics~\cite{Wiseman_Milburn_ch4}) and $P_\text{tot}$ is the total photon flux.
The identification of a work-like part can be justified in different ways.
Intuitively speaking, work is performed on a medium when energy is transferred into it without additional fluctuations.
This should be associated with the emission of coherent light, as would be the case for an ideal laser, which the has minimal possible simultaneous fluctuations in intensity and phase.
Supporting this view, it has been shown that, in a model with coherent driving (see Sec.~\ref{sec:stimulated}), the rate of work output is proportional to the weight of the delta-function peak in $S(\omega)$ at the lasing frequency $\omega_l$~\cite{monsel_energetic_2020,Prasad2024Closing}.
In the dissipative load model considered here, there is no delta-peak, so we instead quantify the work-like quality of the output by the value $S(\omega_l)$.
Due to Eq.~\eqref{eq:power_fraction}, the integral of $S(\omega)$ over a narrow frequency window around $\omega_l$ can therefore be interpreted as the fraction of power that is emitted into that window -- using the fact that the linewidth is much less than $\omega_l$, so all relevant modes have approximately the same energy.

Given an atom-field coupling of the form $W_+ a + W_- a^\dagger$, the input-output formalism implies $G^{(1)}(t,t') \propto \ev{W_+(t) W_-(t')}$, which can be calculated from the quantum regression theorem~\cite{carmichael1993open} in the steady state as
\begin{equation}
    G^{(1)}(t,t')\propto \tr\{W_-(0)e^{(g_u\mathcal{L}_{U}+g_v\mathcal{L}_{V})(t'-t)}[\rho_\infty W_+(0)]\}.
\end{equation}
This, as well as an expression for the second-order correlator $G^{(2)}(t,t')\propto\langle W_+(t)W_+(t')W_-(t')W_-(t)\rangle$, is derived in App.~\ref{app:regression} (see also Chapter 3 of Ref.~\cite{carmichael1993open}).

\begin{figure}
\centering
   \includegraphics[width=0.45\textwidth]{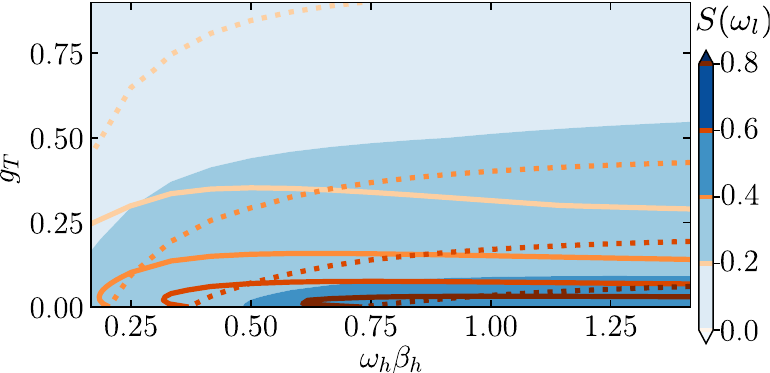}
    \caption{The peak of the intensity spectrum $S(\omega_l)$ of the light emitted into the zero-temperature reservoir is higher for a smaller coupling strength $g_T$. This is true for both the fully symmetric subspace, $(4,0)$ shown by the orange contours as well as the $(2,1)$ irrep shown by blue shaded regions. This is compared to distinguishable particles (dotted contours). Couplings to the hot and cold reservoirs are both set to 0.1 and the cold bath is fixed to $\omega_c=2/3$ and $\beta_c=1.1$. The frequency of the hot bath is $\omega_h=5/3$. }  \label{Fig:intensity}
\end{figure} 

In Fig.~\ref{Fig:intensity}, the intensity spectrum is compared for varying couplings to the dissipative load and temperatures of the hot bath. One can see that, although the energy emitted is lower for a smaller coupling, the intensity spectrum is more sharply peaked -- which suggests the energy is more work-like. If the coupling to the load is fixed we observe that there is an optimal hot bath temperature where coherence is maximised. 

Next, we also study the intensity fluctuations of the emitted light using the normalised second-order correlator~\cite{carmichael1993open}, which for the steady state is
\begin{equation}
    g^{(2)}(t)=\frac{G^{(2)}(t,0)}{\langle W_+W_-\rangle^2}.
\end{equation} 
In particular, we calculate $g^{(2)}(0)$ to characterise these fluctuations. In ideal coherent laser light, one has $g^{(2)}(0)=1$; a value close to one is then an indication that the energy emitted by the system is work-like in the sense of being useful for applications requiring coherent light~\cite{GerryKnightCh5}. 
\begin{figure}
\centering
   \includegraphics[width=0.45\textwidth]{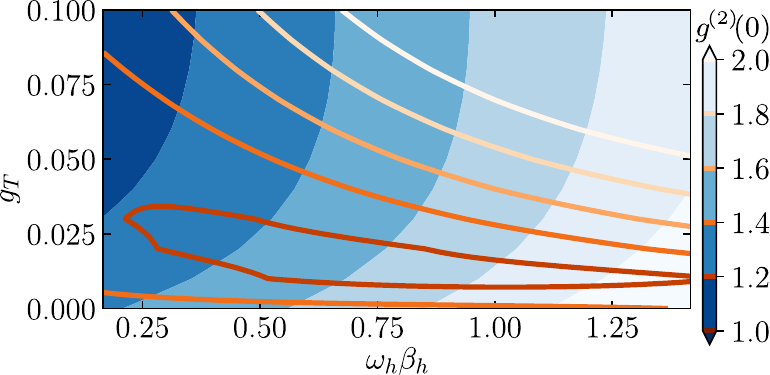}
    \caption{The second-order correlator $g^{(2)}(0)$ is plotted here, showing that the range of parameters for which the statistics of the emitted light are more laser-like (closer to the Poissonian value $g^{(2)}(0)=1$) is quite different for different irreps. However, laser-like behaviour occurs only for small couplings to the dissipative load. Here, $g^{(2)}(0)$ of the fully symmetric subspace $(4,0)$ is shown by the orange contours and by shaded regions for the $(2,1)$ irrep. The cold reservoir is set to $\omega_c=2/3$ and $\beta_c=1.1$ and the frequency of the hot bath is $\omega_h=5/3$.  }  \label{Fig:statistics}
\end{figure}
Fig.~\ref{Fig:statistics} shows that, given the system couples weakly to the load -- which we have already established is required for the emitted energy to be more work-like -- the range of parameters for which $g^{(2)}(0)$ is close to one is quite different for different irreps. We also see that the range of coupling strengths for which coherent light is observed is fairly narrow, regardless of the irrep the system is initialised in. As expected from the intensity spectrum, when this coupling is increased the statistics of the emitted light are more characteristic of thermal light and more characteristic of coherent light for smaller $\beta_h$. Thus, even if the energy current into the zero temperature reservoir is larger, the energy emitted resembles heat and one cannot reasonably conclude that the engine is more powerful. Furthermore, for $(p,q)=(1,0)$ (whose behaviour is equivalent to a single particle),  $g^{(2)}(0)=0$ since two photons cannot be emitted simultaneously.

\subsection{Work from stimulated emission} \label{sec:stimulated}

Finally, we consider the steady-state performance of the ensemble under periodic driving with a resonant classical field~\cite{geva_three-level_1994}. This allows us to properly define the heat current into the system, as well as the power output in the $\omega_l$ transitions. The system Hamiltonian with resonant driving is
\begin{equation}
    {H}_{S}(t)=H_0+\alpha(e^{-i\omega_lt}{W}_++e^{i\omega_lt}{W}_-),
\end{equation} 
where $H_0$ is the free Hamiltonian from Eq.~\ref{eq:free_hamiltonian}. Under weak driving it is possible to model the state evolution using the local Lindblad master equation used in previous sections by simply adding the term obtained from the interaction (driving) Hamiltonian. There may be a concern that this will lead to thermodynamic inconsistency \cite{geva_three-level_1994} because the driving Hamiltonian changes the Bohr frequencies of the system.  However, for weak resonant driving the Floquet approach, which would take these changed Bohr frequencies into account, can lead to an incorrect description because the secular approximation fails~\cite{kolisnyk2024floquet}. In this work, we look only at weak resonant driving and in this case the local master equation is expected to model the dynamics correctly. The master equation
\begin{equation}\label{eq:rotatingframe_masterEq}
     \dot \rho_R=-i[H_{R},\rho_R]+g_u\mathcal{L}_U(\bar{n}_h)\rho_R+g_v\mathcal{L}_V(\bar{n}_c)\rho_R
\end{equation}
is derived in the rotating frame, with interaction Hamiltonian in this frame $H_{R}=\alpha(W_++W_-)$. 
The rate of change in internal energy of the system in the lab frame is \cite{alicki_quantum_1979}
\begin{equation}
   \dot E=\tr \{\dot \rho_S(t) H_S(t)\}+\Tr\{ \rho_S(t) \dot H_S(t)\}.
\end{equation}
The first term defines the total heat current into the system~\cite{monsel_energetic_2020,Prasad2024Closing}.
The second term is identified with the power output of the system. The time dependence cancels in the steady state due to the cyclic properties of the trace, 
\begin{align}
    \mathcal{P}&=-\alpha\omega_l\tr\{({W}_--{W}_+)\rho_R \}
\end{align}
The calculation of the heat currents from the hot and cold baths is done using the adjoint Lindblad operators acting on the Hamiltonian~\cite{kosloff_quantum_2014}, 
\begin{align}
   \mathcal{I}(t)&=\tr\{\rho_S(t) g_u\mathcal{L}_U^{*}(H_S(t))\}+\tr\{ \rho_S(t) g_v\mathcal{L}_V^{*}(H_S(t))\}\\\nonumber
   &=\mathcal{I}_h(t)+\mathcal{I}_c(t).
\end{align}

In the steady state, the efficiency is again independent of the reservoir temperatures and given by $1-\omega_c/\omega_h$ for all irreps. This follows from the same argument given in the non-driven case. That is, in the rotating frame, the time derivative of the operator counting the total number of particles in the ground state, $N_{g}=n/3 + Y/2 - W_z$ must be zero. Therefore, 
\begin{align}
    \partial_t \ev{N_{g}}  =& \langle g_u[(\bar{n}_h+1)U_-U_+-\bar{n}_h U_+U_-]\\\nonumber
    &+\alpha(W_--W_+)\rangle\\\nonumber
    =& \frac{\mathcal{I}_h}{\omega_h} - \frac{\mathcal{P}}{\omega_l}
\end{align}
\begin{figure}
\centering
   \includegraphics[width=0.45\textwidth]{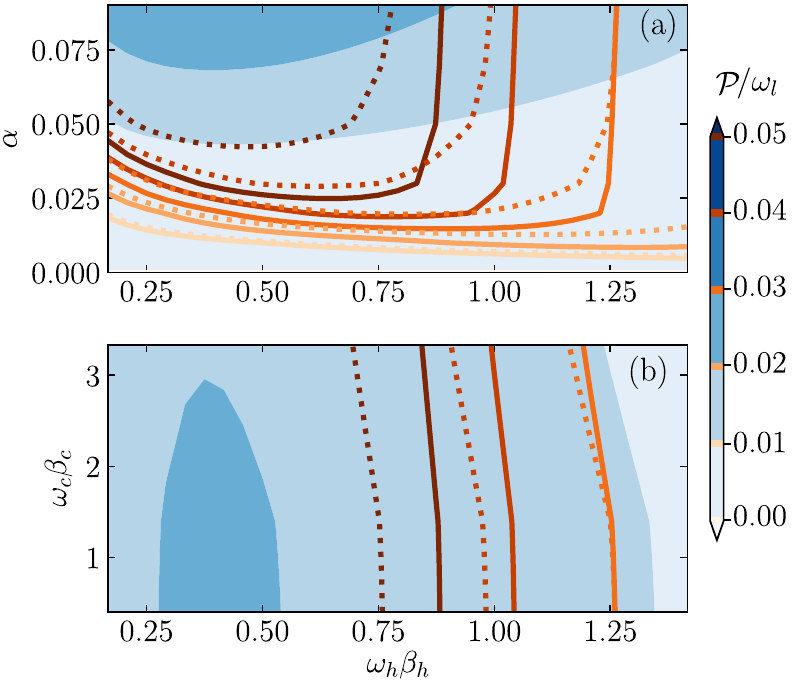}
    \caption{The power emitted by the fully symmetric (4,0) (solid orange contours) and $(2,1)$ irreps (shaded regions) of an ensemble of four, three-level systems is compared the the power emitted by four distinguishable particles (dotted orange contours). The systems are resonantly driven with driving strength $\alpha$ and emitted power increases with driving strength. In (a) this is shown for a cold reservoir set to $\omega_c=2/3$, $\beta_c=1.1$ and $\omega_h=5/3$. In a similar way to the undriven system, the maximum power is not found for the minimum cold reservoir temperature: in (b) the reservoir frequencies are the same as in (a) but the driving strength is fixed to $\alpha=0.07$. The coupling to the hot and cold reservoirs is fixed to 0.1. }  \label{Fig:stim_emission}
\end{figure}
Hence, instead of considering the ratio $\mathcal{P}/\mathcal{I}_h$, we once again quantify the performance of the engine by the absolute power output. The performance is expected to improve with increased driving strength; however, for very strong driving the output driving should drop off since the driving will dominate the dynamics. In Fig.~\ref{Fig:stim_emission}, we show that the fully symmetric state performs better compared to the other irreps, as expected from the results of the previous sections. We also show that four independent systems perform worse than a fully symmetric collective system. Although the emitted power is closer in magnitude to the $(2,1)$ irrep than it was for the system coupled to the dissipative load, it still outperforms the other irreps in this case. We see also that the performance is enhanced when the cold bath is not too cold compared to the hot bath. For $\omega_h\beta_h \ll 1$, the driving will not contribute significantly to the dynamics and hence very little energy can be extracted. Therefore, there is an optimal driving strength depending on the temperature of the hot and cold reservoirs -- which would require a description of strong driving to explore. The description of the system under strong driving goes beyond the scope of this work as the correct description of this limit would require using either the Redfield or Floquet master equations. 

\section{Conclusions \& outlook}
\label{sec:conclusions}
We have analysed the performance of a heat engine made up of a collectively coupled ensemble of three-level systems. Using and extending the representation theoretic tools laid out in Ref.~\cite{yadin_thermodynamics_2023}, we were able to provide a description of these engines going beyond states with bosonic exchange symmetry. This was facilitated by a description of the dynamics with operators acting on the reduced Schur Basis. 
Our first result was to show that a system prepared in a state that is as close as possible to the fully symmetric state will achieve the greatest population inversion when coupled to two heat baths and will thus allow for the highest unitary energy extraction. This was shown both analytically for the limiting case of $\omega_c \beta_c \gg \omega_h \beta_h$ and numerically for a situation where the cold and hot reservoirs are at similar temperatures.
Although the optimality of the fully symmetric subspace here somewhat resembles superradiance, there is no quadratic enhancement of the population inversion in $n$ and so it may be better interpreted as constrained behaviour in the other symmetry subspaces. 
Furthermore, we found that work can be extracted from collective systems outside of the lasing regime of single particles; however, independent particles may outperform the collective system for certain choices of system parameters. 
Next, we showed that the fully symmetric state also produces the highest power output when a dissipative load is weakly coupled to the lasing transition. However, if the load is coupled too strongly, this leads to the emitted light resembling thermal light in intensity fluctuations and with a wider bandwidth, and therefore this energy current resembles heat more than work. Finally, we highlight the differences in performance for different symmetry types for a driven system where now, the work output is clearly defined. In this case we look only at weak driving so that the results are thermodynamically consistent without resorting to Floquet dynamics. 

Although we were able to analytically prove certain properties of the steady states of these systems, some open questions remain.
For example, why does the single-particle population inversion condition apparently extend to a general condition for non-zero lasing ergotropy?
This ergotropy is extracted via processes that move horizontally on the weight diagram, preserving the number of particles in the $\ket{g},\ket{e}$ subspace and moreover the $W$ spin quantum number.
However, it would be interesting to know whether lasing ergotropy always extracts positive work from all such $W$ subspaces simultaneously; this depends on how their populations are ordered from left to right on the diagram.

An immediate future direction would be to study the effect of interactions on the thermodynamics of such systems.
For strong interactions or large particle numbers, one would need a Redfield equation approach due to a break-down of the secular approximation.
Additionally, interactions may allow the weight-space degeneracies in the mixed symmetry subspaces to be lifted. This may also lead to different behaviour to that which is found in this work. 
Also note that our methods are in principle easily extended to higher-dimensional particles, allowing for the study of four-level laser models and other multi-level collective systems.
Another extension is to consider permutation symmetry which is no longer strong in the sense that the symmetry operator commutes with all jump operators, but rather weak in that it commutes only with the Lindbladian~\cite{lieu_symmetry_2020}.
One could also study off-resonant driving and continuous engine cycles with time-varying classical fields~\cite{niedenzu_cooperative_2018}.
Further applications may be found in thermometry, for example, where coupling to two heath baths might allow for simultaneous estimation of two temperatures.

\acknowledgments
We are grateful to Kay Brandner and Sai Vinjanampathy for fruitful discussions.
This project has received funding from the European Union's Horizon 2020 research and innovation programme under the Marie Sk\l odowska-Curie Grant Agreement No.~945422 and the House of Young Talents of the University of Siegen.

%

\appendix
\onecolumngrid

\section{$\mathrm{SU}(3)$ basis}
\label{app:SU3}
We give all commutation relations of the $\mathrm{SU}(3)$ generators for reference~\cite{greiner_quantum_1994}:
\begin{alignat*}{3}
    [W_z, W_\pm] & = \pm W_\pm              & \qquad [W_+, W_-] & = 2W_z \\
    [W_z, U_\pm] & = \mp \frac{1}{2} U_\pm  & \qquad [U_+, U_-] & = \frac{3}{2}Y - W_z \\
    [W_z, V_\pm] & = \pm \frac{1}{2}V_\pm   & \qquad [V_+, V_-] & = \frac{3}{2}Y + W_z \\
    [Y, U_\pm] & = \pm U_\pm                & \qquad [Y, V_\pm] & = \pm V_\pm \\
    [W_+, V_-] & = -U_-                     & \qquad [W_+, U_+] & = V_+ \\
    [U_+, V_-] & = W_- ,                    & &
\end{alignat*}
with the remaining independent ones all vanishing.\\

The generators act on the basis in a $(p,q)$ irrep in the following way~\cite{de_swart_octet_1963}:
\begin{align}
   {W_z}\ket{(p,q),W,w,y}=&t \ket{(p,q),W,w,y} \\\nonumber
   {W_+}\ket{(p,q),W,w,y}=&\sqrt{(W-w) (W+w+ 1)} \ket{(p,q),W,w+1,y} \\\nonumber
   {W_-}\ket{(p,q),W,w,y}=&\sqrt{(W+w) (W-w+ 1)} \ket{(p,q),W,w-1,y} \\\nonumber
   {Y}\ket{(p,q),W,w,y}=&y \ket{(p,q),W,w,y} \\\nonumber
   {V_+}\ket{(p,q),W,w,y}=&A^{p,q}_{W,w,y}\ket{(p,q),W+1/2,w+1/2,y+1}\\\nonumber
   &+ B^{p,q}_{W,w,y}\ket{(p,q),W-1/2,w+1/2,y+1} \\\nonumber
   {V_-}\ket{(p,q),W,w,y}=&A^{p,q}_{W-1/2,w-1/2,y-1}\ket{(p,q),W+1/2,w-1/2,y-1} \\\nonumber
   &+B^{p,q}_{W+1/2,w-1/2,y-1}\ket{(p,q),W-1/2,w-1/2,y-1} \\\nonumber
   {U_+}\ket{(p,q),W,w,y}=&(A^{p,q}_{W,w,y}\sqrt{(W+w+1)(W-w+1)}-A^{p,q}_{W,w-1,y}\sqrt{(W+w)(W-w+1)}) \ket{(p,q),W+1/2,w-1/2,y+1}\\\nonumber
   &+(B^{p,q}_{W,w,y}\sqrt{(W+w)(W-w)}-B^{p,q}_{W,w-1,y}\sqrt{(W+w)(W-w+1)})\ket{(p,q),W-1/2,w-1/2,y+1} \\\nonumber
   {U_-}\ket{(p,q),W,w,y}=&(A^{p,q}_{W-1/2,w-1/2,y-1}\sqrt{(W+w)(W-w)}+A^{p,q}_{W-1/2,w+1/2,y-1}\sqrt{(W+w+1)(W-t)})\\\nonumber
   &\ket{(p,q),W+1/2,w+1/2,y-1} \\\nonumber
   +&(B^{p,q}_{W+1/2,w+1/2,y-1}\sqrt{(W+w+1)(W-w)}-B^{p,q}_{W+1/2,w-1/2,y-1}\sqrt{(W+w+1)(W-w+1)})\\\nonumber
   &\ket{(p,q),W-1/2,w+1/2,y-1} . \\\nonumber
\end{align}
Here, the coefficients $A^{p,q}_{W,w,y}$ and $B^{p,q}_{W,w,y}$ are positive and given by
\begin{equation}
    A^{p,q}_{W,w,y}=\bigg[\frac{(W+w+1)(W(p-q)/3+Y/2+1)((p+2q)/3+W+Y/2+2)((2p+q)/3-W-Y/2)}{2(W+1)(2W+1)}\bigg]^{1/2}
\end{equation}
\begin{equation}
    B^{p,q}_{W,w,y}=\bigg[\frac{(W-w)(W(q-p)/3+W-Y/2)((p+2q)/3-W+Y/2+1)((2p+q)/3+W-Y/2+1)}{2W(2W+1)}\bigg]^{1/2}
\end{equation}.

\section{Constructing the Schur basis}\label{app:young_symmetrizer}
Here we detail the way that the Schur basis described in Sec.~\ref{sec:schur_basis} can be constructed so that operators can be projected onto the reduced basis of a particular irrep of SU(3)~\cite{harrow_applications_2005,Harrow2006}.   
\begin{enumerate}
    \item First, form a partition of $n$ into 3 non-negative integers. These partitions label the irreps of $\mathrm{SU}(3)$. Each partition is represented by a Young diagram labelled by $\lambda_i$ with $n$ boxes and at most $3$ rows. For example, the Young diagrams for $n=3$ are \ytableausetup{smalltableaux} $\lambda_a=$\ydiagram{0+3}, $\lambda_b=$\ydiagram{0+2,1}, $\lambda_c=$\ydiagram{0+1,1,1}.
    
    These correspond to (3,0), (1,1) and (0,0) in the $(p,q)$ notation.
    \item Compute each standard Young tableau $T$ by filling the boxes with $n$ different integers strictly increasing along both the rows and columns. For each $\lambda$, the number of possible standard tableaux is the dimension of the irrep of $\mathrm{S}_n$.
    In our example,  $\lambda_b$ above admits two standard Young tableaux, 
    \ytableausetup{centertableaux}
    \begin{ytableau}
    1 & 2  \\
    3  
    \end{ytableau} and 
    \begin{ytableau}
    1 & 3  \\
    2
    \end{ytableau}
    and so $m_{\lambda_b}=2$. We now fix the tableau $T$ because we will only need to work in the reduced Hilbert space where all the permutation subspaces are traced out. We now also know that the block describing this irrep will appear twice in the block diagonalised Hilbert Space.

\item Fix $T$ to be the standard Young tableau with all numbers in numerical order, e.g.~\ytableausetup{centertableaux}
    \begin{ytableau}
    1 & 2  \\
    3  
    \end{ytableau}. From this standard Young tableau find the permutation operators $P(r)$ and $P(c)$ that permute the numbers along rows and columns. In the $\lambda_b$ example, $P(c)=\{\mathbbm{1},P_{(213)}\}$ and $P(r)=\{\mathbbm{1},P_{(321)}\}$. 

\item Now construct the Young symmetriser, that is the projector \begin{equation}
    \mathcal{P}_{\lambda, T}\propto \left( \sum_{c\in \text{Col}(T)}\text{sgn}(c)P(c)\right) \left(\sum_{r\in \text{Row}(T)}P(r)\right).
\end{equation}
Here, $\text{Col}(T)$ are the permutations of the integers within each column of the diagram and $\text{Row}(T)$ are the permutations of the integers in each row. $\mathcal{P}_{\lambda,T}$ acts on elements of the computational basis to transform them into the block diagonal structure we want, that is, $\mathcal{P}_{\lambda, T}\ket{x_1,x_2...x_n}\propto \ket{\lambda, s_{\lambda}, o_{\lambda}}$. In the $\lambda_b$ example, $\mathcal{P}_{\lambda, T} \propto (\mathbbm{1}+P_{(213)})(\mathbbm{1}-P_{(321)})$ 

\item  Compute the semi-standard tableaux. Informally each box labels a particle which can take one of $d$ computational basis states. Thus the tableau is filled so that numbers do not decrease along rows but strictly increase along columns. In the example of $\lambda_b$, we would get \begin{ytableau}
    0 & 0  \\
    1  
    \end{ytableau},
    \begin{ytableau}
    0 & 0  \\
    2  
    \end{ytableau},
    \begin{ytableau}
    0 & 1  \\
    1  
    \end{ytableau},
    \begin{ytableau}
    0 & 1  \\
    2  
    \end{ytableau},
       \begin{ytableau}
    0 & 2  \\
    1  
    \end{ytableau},
    \begin{ytableau}
    0 & 2  \\
    2  
    \end{ytableau},
    \begin{ytableau}
    1 & 1  \\
    2  
    \end{ytableau} and
    \begin{ytableau}
    1 & 2  \\
    2  
    \end{ytableau}.

\item Construct a type vector $\mathbf{t}(\mathbf{x})$ which counts the number of particles in each of the computational basis states. For example the tableau  \begin{ytableau}
    0 & 0  \\
    1  
    \end{ytableau} has type vector $\mathbf{t}=(2,1,0)$ which has the span $\{ \ket{001}, \ket{010}, \ket{100}\}$ in the computational product basis.       

\item  Construct the restrictions of the permutation operators  $P^{(\mathbf{t})}(r)$ and $P^{(\mathbf{t})}(c)$ onto the subspace of the type vector, e.g.~$P^{(\mathbf{t})}_{(213)}=\ketbra{001}{001}+\ketbra{010}{100}+\ketbra{100}{010}$ for the type vector $\mathbf{t}=(2,1,0)$, and then calculate the Young symmetriser. Note that semi-standard tableaux with the same numbers will lead to the same type vectors, e.g. \begin{ytableau}
    0 & 1  \\
    2  
    \end{ytableau}and
       \begin{ytableau}
    0 & 2  \\
    1  
    \end{ytableau} are both spanned by $\{ \ket{012}, \ket{021}, \ket{210}\}$ however, they have different Young symmetrisers. In the Schur Basis discussed in Sec.~\ref{sec:schur_basis}, the states corresponding to these tableaux would have the same weights $(w,y)$. 
\item Once the projector is constructed, perform a singular value decomposition on it. The first $k$ columns of the left singular vector with non-zero singular values form the basis elements $\ket{\lambda, o_{\lambda}}$ 
\item Operators that were originally expressed in the Hilbert space $\mathcal{H}^n$ can now be projected onto this reduced basis by, $O_{o,o'}^{\lambda}=\bra{\lambda, o_{\lambda}}O_n\ket{\lambda, o_{\lambda}'}$. 
\end{enumerate}

\section{Structure of the steady state}\label{app:steady_state}

Here, we show that the steady state of the master equation \eqref{eq:ergotropy_master} has a block-diagonal structure with respect to the Schur basis.
In fact, we can assume more generally that the master equation contains the dissipators $\mathcal{D}[L_i]$ for $L_i \in \{W_\pm,\, U_\pm,\, V_\pm\}$, with the Hamiltonian being a linear combination of the diagonal generators $W_z,\, Y$.
All the jump operators are then generalised ladder operators for each of $W_z$ and $Y$, as seen from the commutation relations in App.~\ref{app:SU3}.

It follows that the master equation generator $\mathcal{L}$ is covariant with respect to diagonal unitary transformations $\mathcal{U}_{\theta,\phi} := e^{-i(\theta W_z + \phi Y)} \cdot e^{i(\theta W_z + \phi Y)}$, i.e., $[\mathcal{L}, \mathcal{U}_{\theta,\phi}] = 0$.

Suppose we are given any steady state $\rho$, so $\mathcal{L}(\rho)$ = 0.
Then we show that it is possible to construct another steady state $\sigma$ that is also invariant under the unitaries $\mathcal{U}_{\theta,\phi}$.
This state is given by ``twirling'' over the diagonal subgroup -- in other words, averaging $\rho$ over all $\theta$ and $\phi$:
\begin{align}
    \sigma := \int_0^{2\pi} \frac{\dd \theta}{2\pi} \int_0^{2\pi} \frac{\dd \phi}{2\pi} \; \mathcal{U}_{\theta,\phi}(\rho).
\end{align}
It is clear that $\sigma$ is invariant under the diagonal subgroup by construction, and
\begin{align}
    \mathcal{L}(\sigma) & = \int_0^{2\pi} \frac{\dd \theta}{2\pi} \int_0^{2\pi} \frac{\dd \phi}{2\pi} \; \mathcal{L} \circ \mathcal{U}_{\theta,\phi}(\rho) \nonumber \\
        & = \int_0^{2\pi} \frac{\dd \theta}{2\pi} \int_0^{2\pi} \frac{\dd \phi}{2\pi} \; \mathcal{U}_{\theta,\phi} \circ \mathcal{L}(\rho) \nonumber \\
        & = 0.
\end{align}
Then we see that $\sigma$ is block-diagonal in the Schur basis with respect to $w$ and $y$, meaning that
\begin{align}
    \mel{\lambda,W,w,y}{\sigma}{\lambda',W',w',y'} = 0 \quad \text{if } \lambda \neq \lambda', \,  w \neq w',\, \text{or } y \neq y'.
\end{align}
(The block-diagonal structure with respect to the irrep $\lambda$ follows from Theorem 1 of Ref.~\cite{yadin_thermodynamics_2023}.)
Note that coherences are guaranteed to vanish only between different weight spaces; we cannot say anything about different $W$ in the same weight space.

For a single component in the irrep space $\mathcal{H}^\lambda$, Theorem 1 of Ref.~\cite{yadin_thermodynamics_2023} tells us that the steady state is typically unique.
This holds up to certain specific fine-tuning cases of the rates in the master equation -- for the master equation considered here, we require only that at least two of the coefficients $g_u,g_v,g_t$ are non-zero.
It then follows that the only steady state $\sigma^\lambda$ has the given block-diagonal structure.

Note that these observations can all be easily generalised to higher-dimensional particles, under the assumption that the jump operators are elements of the Cartan basis, and the Hamiltonian is a linear combination of the diagonal Cartan basis elements.

\section{Ergotropy in the low-temperature cold bath limit}\label{app:cold}

Here we show that in the limiting case of $\omega_c\beta_c \gg 1$, it is possible to find the solution for the steady state of the model with no lasing transition, and determine the extractable energy in a given irrep.\\

In the case that $\omega_c\beta_c \gg 1$, the average occupation number of the cold bath is effectively zero. Therefore, the master equation Eq.~\ref{eq:ergotropy_master} becomes
\begin{equation}
    \dot{\rho}=g_v({V_+}\rho{V_-}-\frac{1}{2}\{{V_-}{V_+},\rho\})+g_u\mathcal{L}_U(\bar{n}_h)\rho
\end{equation}

\begin{figure}[h]
    \centering
       \includegraphics[width=0.35\textwidth]{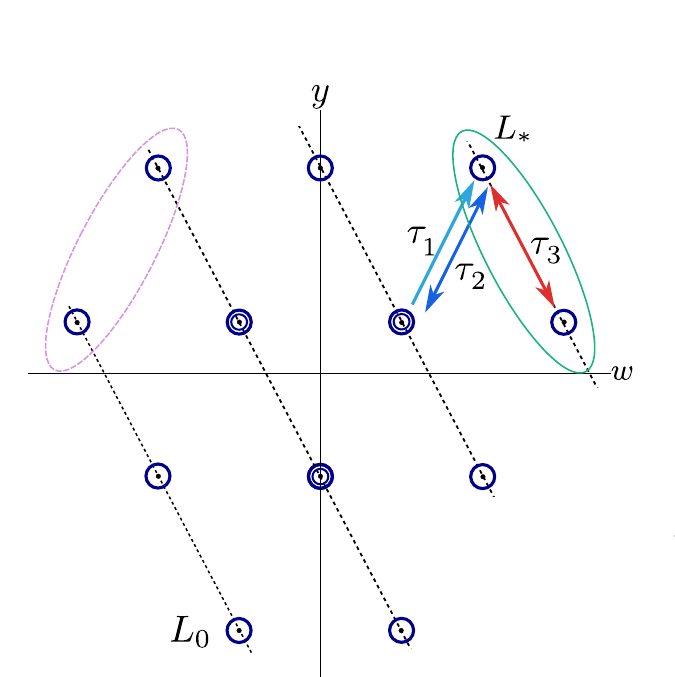}
        \caption{A diagram of the reduced basis states $\ket{(p,q),W_i,w_i,y_i}$ of one of the irreducible representations found for four 3-level systems, labelled $\lambda = (p,q) =(2,1)$. Dotted lines form the family of lines with equation $L=y/2+w$. The arrows labelled by $\tau_i$ depict the action of each of the terms in Eq~\eqref{eq:zero_Tterms}.
        The points inside the green ellipse form the line $L_*$ on which the steady state lives for a low-temperature cold bath. The reflected steady state $\rho_r$ is made up of the populations inside the purple dashed ellipse.}  \label{Fig:basis_app}
    \end{figure}
In the space of basis vectors, define the family of lines indexed by the numbers $L=y/2+w$ which can be traversed within the diagram by applying $U_{\pm}$. 

Fig.~\ref{Fig:basis_app} illustrates the lines that are present for the $(p,q)=(2,1)$ irrep, which is a possible irrep if there are 4 particles in the ensemble. Consider the sum of the populations along one of these lines which is constant in the steady state \cite{latune_thermodynamics_2019}, 
\begin{equation}
 \sum_{i \in L}\dot{\rho}^{i}_L:=\sum_{w_i+y_i/2=L}\bra{p,q,W_i,w_i,y_i}\dot{\rho}\ket{(p,q),W_i,w_i,y_i}=0.  
\end{equation}
Where $i\in L$ labels each of the Schur Basis states that are possible for the irrep labelled $(p,q)$ on a particular line $L$.
In order to get to an expression for the steady state, we break up the total sum into three parts 
\begin{align}\label{eq:zero_Tterms}
 \sum_{i \in L}\dot{\rho}^{i}_L&=\tau_1^{(L)}+\tau_2^{(L)}+\tau_3^{(L)}\\\nonumber
 &=g_v\sum_{i \in L}\bra{(p,q),W_i,w_i,y_i}{V_+}\rho{V_-}\ket{(p,q),W_i,w_i,y_i}\\\nonumber 
 &-\frac{g_v}{2}\sum_{i \in L}\bra{(p,q),W_i,w_i,y_i}\{{V_-}{V_+},\rho\}\ket{(p,q),W_i,w_i,y_i}\\\nonumber
 &+g_u\sum_{i \in L}\bra{(p,q),W_i,w_i,y_i}\mathcal{L}_U\rho\ket{(p,q),W_i,w_i,y_i}.  
\end{align}
Since the populations are constant for each single line,
$\tau_3^{(L)}=0$ for all $L$. 

Starting from the line that is furthest on the left in Fig.~\ref{Fig:basis_app}, $L_0=(p-2q)/3$, $\tau_1^{(L_0)}$
vanishes since there are no states on the line $L_{(-1)}=L_0-1$. Then because we consider only the steady state, due to Eq.~\eqref{eq:zero_Tterms}, $\tau_2^{(L_0)}=0$. Then, using the operator coefficients from App.~\ref{app:SU3},
\begin{align}
    {V_-}{V_+}\ket{(p,q),W,w,y}=&(B^{p,q}_{W+1,w,y}A^{p,q}_{W,w,y}+A^{p,q}_{W-1,w,y}B^{p,q}_{W,w,y})\ket{(p,q),W,w,y}\\\nonumber
    &+(A^{p,q}_{W,w,y})^2\ket{(p,q),W+1,w,y}+(B^{p,q}_{W,w,y})^2\ket{(p,q),W-1,w,y}
\end{align}
the expectation value in each term of $\tau_2$ is always non-negative, since all the coefficients are non-negative.

Therefore, for the sum of all the terms to be zero, the total population at each point on the line must be zero. We can then proceed through each of the lines in the irrep as follows: $\tau_1^{(L+1)}=0$ since it depends on the populations from the previous line which are all zero, thus once again we are left with $\tau_2^{(L+1)}=0$ and the populations along this line are all individually zero by the same argument as before. This is extended until the last possible line in the state space, $L_* = (2p+q)/3$: in this line $\tau_2^{(L_*)}$ 
automatically vanishes since there are no ${V_+}$ transitions allowed. Then the populations from the previous line are all zero, which makes $\tau_1^{(L_*)}=0$. Thus, we are left with $\tau_3^{(L_*)}=0$. The individual populations on this line will have a Boltzmann distribution since they are determined by dynamics with a single heat bath, 
\begin{equation}
  \dot{\rho}_{2p+q}=\mathcal{L}_U\rho_{2p+q},
\end{equation}
and futhermore are non-degenerate since they lie on the boundary of the weight diagram~\cite{greiner_quantum_1994}.
The magnitude of the populations along this line decrease with decreasing $y$ coordinate.\\

The lasing ergotropy is found from Result~\ref{res:ergotropy} (a) with the Boltzmann distribution over the upper-right line, using the fact that the points on this line have coordinates $w_k = (p+k)/2$, $y_k = \text{const.} - k$ for $k = 0,1,\dots,q$:
\begin{align}
    \mathcal{E}_l & = 2 \omega_l \tr[W_z \rho_\infty^\lambda] \nonumber \\
        & = \omega_l \sum_{k=0}^q \frac{e^{-\beta_h \omega_h}}{Z} (p + k),
\end{align}
with $Z = \sum_{k=0}^q e^{-\beta_h \omega_h}$.
Performing the sum, we obtain
\begin{align}
    \mathcal{E}_l & = \omega_l \left[ p + q + \frac{1}{1-e^{-\beta_h \omega_c}} - \frac{q+1}{1-e^{-(q+1)\beta_h \omega_h}} \right].
\end{align}\\

We can also make statements about the full ergotropy in this limit.
In particular, the full ergotropy $\mathcal{E}$ equals the lasing ergotropy $\mathcal{E}_l$ if the state obtained by extracting $\mathcal{E}_l$ is also a passive state -- i.e., when no more energy could be extracted by any unitary operation acting on the irrep.
This state, denoted $\rho_r$, is the reflection of the one described above in the $w$-axis on the weight diagram.
While the steady state lives on the line $L_*$, $\rho_r$ lives on its reflection $L_r$.
Since the energies at points on $L_r$ increase as $y$ decreases (moving down the diagram), and the populations also decrease by the Boltzmann ratio $e^{-\beta_h \omega_h}$, we see that $\rho_r$ is passive whenever all the points lying on $L_r$ are the $(q+1)$ lowest energy eigenstates.

The ground state is the upper-left point in the diagram.
Relative to this, the energies of the points on $L_r$ are $\omega_c, 2\omega_c, \dots, q \omega_c$ (moving downwards, or applying $V_-$).
Considering moving in the two other directions from the ground state, applying $W_+$ adds energy $\omega_l$, while $U_-$ adds $\omega_h > \omega_l$.
So the lowest $(q+1)$ energies must lie on $L_r$ whenever $q \omega_c \leq \omega_l$.
This is then the condition for the equality $\mathcal{E} = \mathcal{E}_l$ for the case of a low-temperature cold bath.

\section{Quantum regression theorem}~\label{app:regression}
Here, we show how two-time correlators like $G^{(1)}(t,t')\propto \expval{W_+(t)W_-(t')}$ and $G^{(2)}(t,t')\propto \expval{W_+(t)W_+(t')W_-(t')W_-(t)}$ can be calculated using the quantum regression theorem~\cite{carmichael1993open}. First let $\sigma$ be the density operator for the entire system and environment state such that $\tr_E(\sigma)=\rho$ and let $H$ be the Hamiltonian acting on this space. Then, we will calculate the second-order correlator for $t<t'$ as this result is easily adapted to give the result of the first-order correlator and these are the times we are focus on in this work. Consider then, the second-order correlator:
\begin{equation}
    \expval{W_+(t)W_+(t')W_-(t')W_-(t)}=\tr_{S}\{W_+(0)W_-(0)\tr_E[e^{-iH(t'-t)}W_-(0)\sigma(0)W_+(0)e^{iH(t'-t)}]\}
\end{equation}
Where we have used that the operators $W_{\pm}(0)$ only act on the system part of the Hilbert space. Then set $\sigma_T(t'-t)=e^{-iH(t'-t)}W_-(0)\sigma(0)W_+(0)e^{iH(t'-t)}$ 
and note that for for system dynamics generically given by $\dot\rho=\mathcal{L}\rho$,
\begin{equation}
    \tr_E[\sigma_T(t'-t)]=e^{\mathcal{L}(t'-t)}(W_-(0)\rho(t)W_+(0)).
\end{equation}
Therefore the correlator becomes, 
\begin{equation}
     \expval{W_+(t)W_+(t')W_-(t')W_-(t)}=\tr_{S}\{W_+(0)W_-(0)e^{\mathcal{L}(t'-t)}(W_-(0)\rho(t)W_+(0))\}.
\end{equation}
The first-order correlator can be calculated in the same way by replacing $W_+(t')W_-(t')$ with $W_-(t')$ and $W_-(t)$ with the identity. 
\end{document}